\documentclass[12pt]{article}
\usepackage{epsfig}
 
\newcommand{\mrm}[1]{\mathrm{#1}}
 
\newcommand{\lessim}{\raisebox{-0.8mm}%
{\hspace{1mm}$\stackrel{<}{\sim}$\hspace{1mm}}}
 
\newcommand{\alphaem}{\alpha_{\mrm{em}}}
 
\renewcommand{\d}{\mrm{d}}
\newcommand{\e}{\mrm{e}}

\newcommand{\g}{\mrm{g}}

\newcommand{\p}{\mrm{p}}
\newcommand{\q}{\mrm{q}}

\renewcommand{\u}{\mrm{u}}

\newcommand{\Z}{\mrm{Z}}

\newcommand{\qbar}{\overline{\mrm{q}}}

{\end{list}}
\newcounter{enumct}

\newlength{\abstwidth}
\newlength{\captivewidth}
\newcommand{\captive}[1]{\rule{5mm}{0mm}%
\begin{minipage}{\captivewidth}%
\caption[small]{#1}\end{minipage}}
 
\begin{document}
 
\sloppy
 
\pagestyle{empty}
 
\begin{flushright}
LU TP 97--04 \\
March 1997
\end{flushright}
 
\vspace{\fill}
 
\begin{center}
{\LARGE\bf QCD aspects of leptoquark production}\\[3mm]
{\LARGE\bf at HERA }\\[10mm]
{\Large C. Friberg\footnote{christer@thep.lu.se}, %
E. Norrbin\footnote{emanuel@thep.lu.se} and %
T. Sj\"ostrand\footnote{torbjorn@thep.lu.se}} \\[3mm]
{\it Department of Theoretical Physics,}\\[1mm]
{\it Lund University, Lund, Sweden}
\end{center}
 
\vspace{\fill}
 
\begin{center}
{\bf Abstract}\\[2ex]
\begin{minipage}{\abstwidth}
If a leptoquark is produced at HERA as a narrow resonance, various effects 
tend to broaden the measurable mass distribution considerably. These effects
are discussed here, with special emphasis on initial- and final-state QCD 
radiation. A proper understanding is important to assess the significance
of data and to devise strategies for better mass reconstruction.
\end{minipage}
\end{center}
 
\vspace{\fill}
 
\clearpage
\pagestyle{plain}
\setcounter{page}{1}

Recently, the H1 and ZEUS Collaborations at HERA have presented evidence for
an excess of events at large $Q^2$ and $x$ \cite{H1, ZEUS}. This could be
nothing but a statistical fluke. Alternatively, it could be the first signal
of the production of an $s$-channel resonance, a leptoquark (LQ). 
In order to test such a hypothesis, as further data accumulate,
it is important to understand the production characteristics of an 
LQ at HERA. In particular, it should be noted that an expected
narrow mass peak will be smeared by various physics and detector effects.
In this letter we want to give a brief survey of the physics components
that could contribute to this smearing, and estimate the magnitude of each of 
these effects. Previous studies of a similar kind have been performed with the
{\sc Lego} generator \cite{LEGO}, which is partly based on {\sc Pythia} 
\cite{Manual}. We also want to give some examples of how to improve the 
reconstruction of the LQ mass.

This is not a study on the physics implications of an LQ observation at HERA
\cite{theorists}. Therefore we do not discuss the origin of an LQ in terms of an 
underlying theory, be that compositeness, supersymmetry or anything else, but 
stay with a purely phenomenological description of LQ properties \cite{LQphen}. 
Even so, several options are possible, and we will restrict ourselves
further. An LQ may have spin 0 or 1, with differences in the decay angular 
distribution, but of little importance for the considerations in this letter,
so we stay with the spin 0 alternative. An LQ may have net fermion number
0 ($\q\overline{\ell}$ or $\qbar\ell$) or $\pm 2$ ($\q\ell$ or 
$\qbar\overline{\ell}$). Given that HERA has been running with an $\e^+$ beam 
in recent years, the latter kind would be disfavoured by requiring a sea 
antiquark from the proton. It would have been favoured in the earlier HERA runs 
with $\e^-$, however, where the valence quark distributions could have been 
accessed. The ratio in parton distributions between valence and sea at 
$x \approx 0.5$ being larger than the recorded luminosity ratio 
$\e^+ \p / \e^-\p$, the previous non-observation \cite{HERAnonobs} favours LQ's  
with vanishing fermion number. A leptogluon scenario is not excluded, but is 
not favoured \cite{leptog}. QCD consequences of a $\qbar\overline{\ell}$ LQ
or a leptogluon are briefly mentioned later.    
 
Phenomenologically, the production cross section for a leptoquark in the process
\mbox{$\q + \ell^+ \to \mrm{LQ}$} can be written as \cite{LQphen}
\begin{equation}
\sigma = \frac{\pi\lambda^2}{4 s} q(x,M_{\mrm{LQ}}^2)
       = k \frac{\pi^2\alphaem}{s} q(x,M_{\mrm{LQ}}^2)  ~.
\label{sigmasimple}
\end{equation}
Here $\lambda$ gives the strength of the unknown Yukawa coupling;
the alternative $k$ parameter is normalized such that $k=1$
corresponds to electromagnetic strength of the coupling.
To first approximation $x = M_{\mrm{LQ}}^2/s$ --- corrections to this will
be a main theme of the letter --- and $q(x,M_{\mrm{LQ}}^2)$ the parton 
distribution at a scale given by the LQ mass.

The observed handful of candidate events per experiments, for an integrated
luminosity of order 15 pb$^{-1}$ and a detection efficiency around 50\%
(including a cut $Q^2 > 15000$~GeV$^2$),
would suggest a production cross section of the order of 1 pb. For a $\u\e^+$ LQ
of around 200 GeV mass this corresponds to $k \sim 0.01$. The width of a scalar LQ,
\begin{equation}
\Gamma_{\mrm{LQ}} = \frac{\lambda^2}{16\pi} M_{\mrm{LQ}} 
       = \frac{k \alphaem}{4} M_{\mrm{LQ}} ~,
\end{equation} 
then becomes only $\Gamma_{\mrm{LQ}} \sim 4$~MeV. A $\d\e^+$ LQ would be somewhat 
broader, but in either case the width is negligibly small on an experimental scale.
Unless the total width is enhanced significantly by decays to exotic channels,
the LQ is long-lived enough that it will form an LQ-hadron, made up of the 
LQ and an antiquark or a diquark. 

QCD radiation of gluons and QED radiation of photons in principle can occur in the
initial state, off the LQ itself, and in the final state. Including all interference
terms, a complicated radiation pattern is then possible. The small width here offers
a considerable simplification: interference terms and radiation off the LQ are
suppressed for radiated energies above $\Gamma_{\mrm{LQ}}$ \cite{Gammaiscut,topdecay}. 
The region $E \lessim \Gamma_{\mrm{LQ}} \sim 10$~MeV giving a very small contribution 
to the total radiated energy, it is therefore sufficient to consider three main classes 
of corrections, (i) initial-state radiation, (ii) LQ-hadron formation, and (iii)
final-state radiation. We will consider these effects one at a time, roughly in order
of decreasing importance.

\begin{figure}[tp]
\begin{center}
\mbox{\epsfig{file=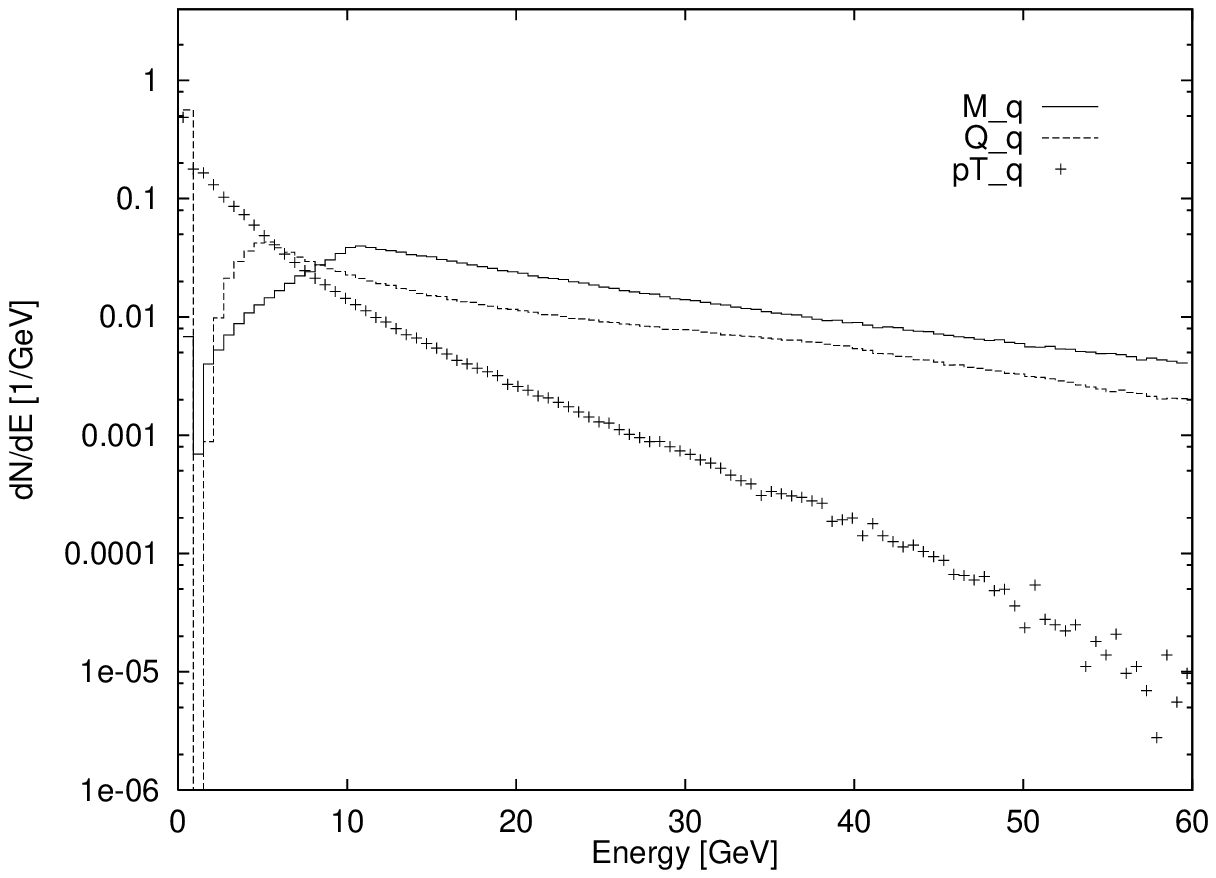}} 
\end{center}
\captive{Distribution of $M_{\q}$ (full), $Q_{\q}$ (dashed) and 
$p_{\perp \q}$ (crosses).
\label{mqptdist}}
\end{figure}

The main effect of final-state radiation is that the hadronic system of the LQ
decay acquires a mass in excess of the naive quark one and that, as a consequence, 
the energy of the recoiling lepton is reduced. The jet mass phenomenology is not very 
different from experience in hadronic $\e^+\e^-$ annihilation events. A standard
parton-shower description \cite{finshow} tuned to LEP data predicts an average quark
jet mass of about 30 GeV at 200 GeV energy. Taking over the same formalism for
the LQ decays then gives an average mass $\langle M_{\q} \rangle \approx 32$~GeV,
Fig.~\ref{mqptdist}, i.e. somewhat higher
since the absence of a radiating final-state partner removes some phase-space 
competition. In a more detailed study one should include the explicit 
matrix-element information for the region of well-separated emission, which could 
introduce some modest dependence on the spin of the leptoquark.

\begin{figure}[tp]
\begin{center}
\mbox{\epsfig{file=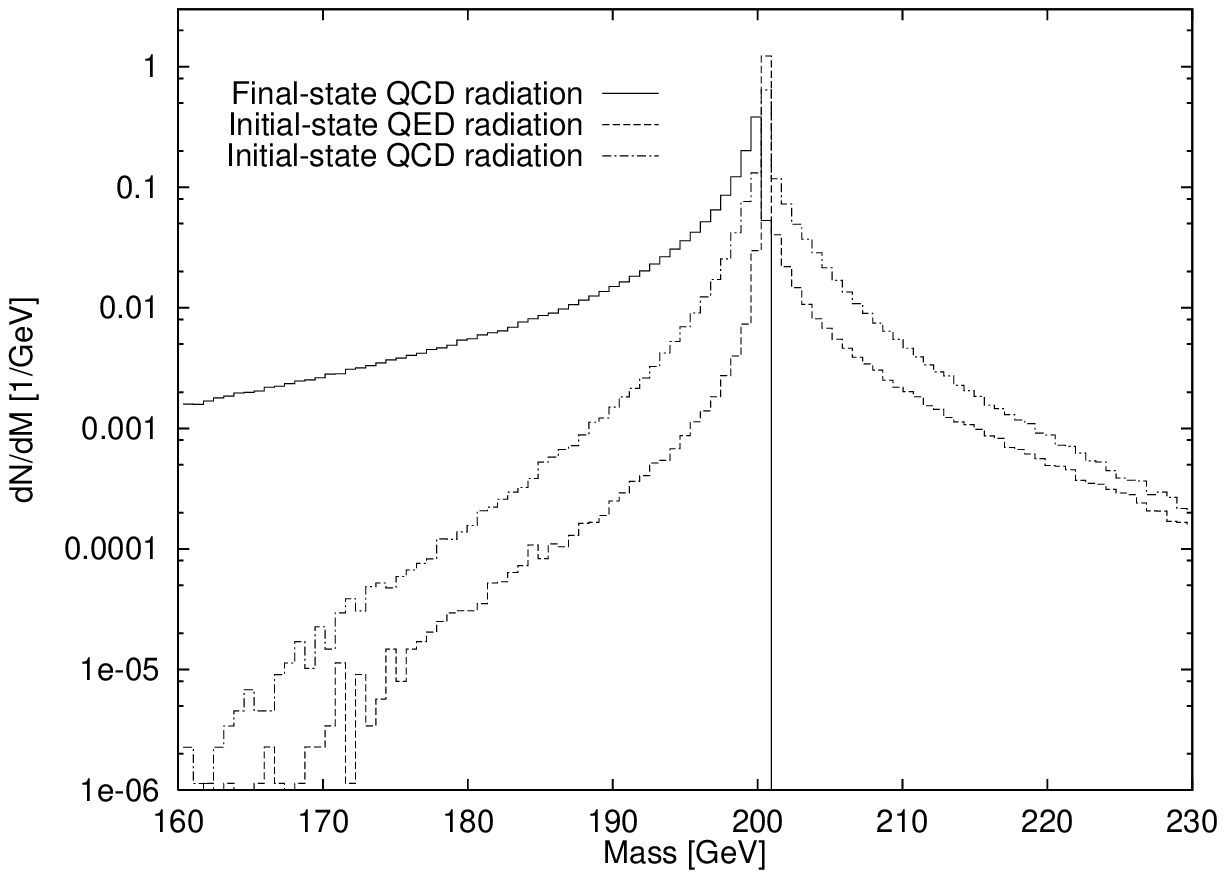}} 
\end{center}
\captive{The reconstructed LQ mass distribution 
$m_{\mrm{recon}} = \sqrt{x_{\mrm{Bj}} s}$ for an input mass of
200 GeV, with a simple cut $Q^2 > 15000$~GeV$^2$ on the leptoquark 
sample. Results are shown with only one component active at a time:
final-state QCD radiation (full), initial-state QED radiation off 
the positron (dashed) and initial-state QCD radiation (dot-dashed).
\label{mlqrecon}}
\end{figure}

In the rest frame of the LQ the lepton takes an energy
\begin{equation} 
E_{\ell} = \frac{M_{\mrm{LQ}}}{2} \left( 1 -  \frac{M_{\q}^2}{M_{\mrm{LQ}}^2} 
\right) ~.
\end{equation}   
The rest-frame lepton scattering angle $\theta^*$ is only 
very little affected by the QCD radiation \cite{Kleiss}. (Furthermore, for a spin 0 
leptoquark, the inclusive decay distribution is isotropic in $\cos\theta^*$ in any 
case.) Even though the standard DIS variables do not have the traditional meaning
related to a spacelike boson propagator, they can be defined purely experimentally.
The $M_{\q}$ term above then propagates to give
\begin{eqnarray}
Q^2 & = & \frac{M_{\mrm{LQ}}^2}{2} \left( 1 - \cos\theta^* \right) 
          \left( 1 -  \frac{M_{\q}^2}{M_{\mrm{LQ}}^2} \right) ~,
    \label{qtwomod}\\
x_{\mrm{Bj}} = \frac{Q^2}{2Pq} & = & \frac{M_{\mrm{LQ}}^2}{s} 
          \frac{Q^2}{Q^2 + M_{\q}^2} ~.
\end{eqnarray}
Defining $\tau$ to be the $x$ value relevant for an LQ mass determination,
\begin{equation}
\tau \equiv \frac{\hat{s}}{s} = \frac{M_{\mrm{LQ}}^2}{s} =
     x_{\mrm{Bj}} \left( 1 +  \frac{M_{\q}^2}{Q^2} \right) ~,
\label{timelikeshift}
\end{equation}
one finds that $\tau > x_{\mrm{Bj}}$. In particular note that the denominator
of the $M_{\q}$ correction factor is $Q^2$, not $M_{\mrm{LQ}}^2$. This means that 
low-$Q^2$ data would not be expected to show any peak in $x_{\mrm{Bj}}$.
A typical cut $Q^2 > 15000$~GeV$^2$ reduces the $\langle M_{\q} \rangle$ from 
32~GeV to 29~GeV (eq.~(\ref{qtwomod})). The 
$\langle M_{\q}^2 \rangle \approx 1430$~GeV$^2$ then gives an average correction 
factor in eq.~(\ref{timelikeshift}) of somewhat above 5\%. If the experimentally
reconstructed LQ mass is taken to be $\sqrt{x_{\mrm{Bj}} s}$, the original 
$\delta$ function is smeared as shown in Fig.~\ref{mlqrecon}. Since 
$Q^2 = x_{\mrm{Bj}} y s$ is an identity that follows from the definition of the
respective variable, the alternative experimental measure $\sqrt{Q^2/y}$
gives the same smearing.

Also photons may be radiated in the final state, both from the lepton and the
quark. Owing to the smaller coupling, the amount of radiation is reduced compared
with the QCD case above. Radiation almost collinear with the lepton occurs at a
significant rate, but is not resolved by the calorimetric definition of lepton
energy and so is of little consequence. In our studies we choose not to resolve
final-state photons below a 1~GeV invariant mass cut-off.

We next turn to initial-state radiation, i.e. radiation off the incoming quark
and lepton lines. This reduces the longitudinal momentum fraction carried by
the reacting quark/lepton, and builds up a $p_{\perp}$ and a spacelike virtuality
for it. 

Photon radiation off the lepton line is dominated by the almost collinear one.
In this limit, one has that $\tau = x_{\mrm{q}} x_{\ell}$, where the $x_i$ are 
the respective momentum fractions. In the leading-log approximation, the
positron-inside-positron distribution is roughly \cite{LEP1QED}
\begin{equation}
D_{\e}^{\e}(x_{\ell},M_{\mrm{LQ}}^2) \approx \beta (1-x_{\ell})^{\beta - 1}%
~~~~\mrm{with}~~~~\beta = \frac{\alphaem}{\pi} \left( 
\ln \frac{M_{\mrm{LQ}}^2}{m_{\e}^2} - 1 \right) \approx 0.0575 ~. 
\end{equation}
The eq.~(\ref{sigmasimple}) is then modified to
\begin{equation}
\sigma = k \frac{\pi^2\alphaem}{s} \int \hspace{-2mm} \int \d x_{\q} \; 
\d x_{\ell} \; q(x_{\q},M_{\mrm{LQ}}^2) \, D_{\e}^{\e}(x_{\ell},M_{\mrm{LQ}}^2) \,
\delta ( x_{\mrm{q}} x_{\ell} - \tau) ~.
\end{equation}
One obtains $\langle x_{\ell} \rangle \approx 0.99$ or 
$\langle x_{\q} \rangle \approx 1.01 \tau$. The steep fall-off of 
$q(x_{\q},M_{\mrm{LQ}}^2)$ dampens the tail to large $1-x_{\ell}$.
Therefore typical experimental cuts on photon energy lost in the beam pipe 
do not make a big difference. The relation between the true 
and reconstructed masses is
\begin{equation}
\tau = x_{\mrm{Bj}} \, \frac{x_{\ell}^2}{1 - (1- x_{\ell}) /y } ~,
\label{photonshift}
\end{equation}
so the mass shift may be in either direction, Fig.~\ref{mlqrecon}. 

Radiation of photons and gluons in the initial state leads to a buildup of 
spacelike virtualities and transverse momenta for the incoming quark and lepton. 
The largest effects come on the quark side, Fig.~\ref{mqptdist}, where 
$\langle Q_{\q} \rangle \approx 14.3$~GeV and
$\langle p_{\perp\q} \rangle \approx 3.6$~GeV, while
$\langle Q_{\ell} \rangle \approx 0.85$~GeV and
$\langle p_{\perp\ell} \rangle \approx 0.15$~GeV for the incoming positron,
using the spacelike parton-shower formalism of \cite{inishow}. (Numbers vary
a bit depending on choice of parton distribution parametrizations etc.)
The spacelike virtualities can be seen as a kinematical consequence of the 
$p_{\perp}$ kicks, so the $Q_{\q(\ell)}$ and $p_{\perp\q(\ell)}$ are 
strongly correlated. Note that
$\langle Q_{\q}^2 \rangle \ll M_{\q}^2$: spacelike parton-shower evolution
is constrained by the limited phase space and the steeply falling parton 
distributions at large $x$. That is, the probability that the daughter quark
at $x_i$ comes from the branching of a mother quark at $x_{i-1} > x_i$ is 
related to the ratio $q(x_{i-1})/q(x_i)$ of parton distributions, and the 
integral of $q(x_{i-1})$ over $x_{i-1} > x_i$ is small. At each branching 
one has $p_{\perp i}^2 \approx (1-z_i) Q_i^2$, with $z_i= x_i / x_{i-1}$, 
so from the argument above it is clear why also
$\langle p_{\perp\q}^2 \rangle \ll \langle Q_{\q}^2 \rangle$.   

The consequences of nonzero virtualities and transverse momenta can again be 
traced through the standard DIS variables. Also including the previous
effects, i.e. $M_{\q} \neq 0$ and $x_{\ell} \neq 0$, gives the relation
\begin{eqnarray}
\tau & = & \frac{ b - 2 a Q_{\q}^2 + \sqrt{b^2 - 4 a x_{\ell} Q^2 \left( %
   Q_{\q}^2 - p_{\perp \q}^2 \right)} }{ 2 s a } ~, \nonumber \\
\mrm{where}~~~a & = & 1 - \frac{1-y}{x_{\ell}} 
   \label{totalshift} \\
\mrm{and}~~~b & = & x_{\ell} Q^2 + M_{\q}^2 + Q_{\q}^2 - 
   2 \overline{p}_{\perp\q } \overline{k}'_{\perp} ~.  \nonumber
\end{eqnarray} 
(A correction formula in terms of another set of variables is found in
\cite{LEGO}.) 
The spacelike virtuality and $p_{\perp}$ kick may either increase or decrease 
the estimated LQ mass. In particular, there is an explicit dependence on the 
azimuthal angle between the $p_{\perp\q}$ of the incoming quark 
($=p_{\perp \mrm{LQ}}$ in this approximation) and the $k'_{\perp}$ of the 
outgoing lepton. The resulting spread in reconstructed LQ mass is shown in 
Fig.~\ref{mlqrecon}.

A $p_{\perp\q}$ contribution is also given by the primordial $k_{\perp}$ of the
parton-shower initiator inside the proton. This primordial $k_{\perp}$ is
some combination of Fermi motion and initial-state radiation below the soft cut-off
of the simulation program. A typical scale for the former would be 
$\langle k_{\perp}^2 \rangle \approx 0.25$~GeV$^2$, but even if this is increased 
to 1~GeV$^2$ the net effects are quite negligible, since the $k_{\perp}$
is added to the much larger $p_{\perp}$ of the perturbative parton shower.
Only in some future study of the LQ $p_{\perp}$ distribution itself would the 
low-$p_{\perp\mrm{LQ}}$ tail be sensitive to the choice.

The effects of LQ-hadron formation are not large, since the LQ is so massive that 
its motion is not significantly affected by hadronization effects.
Data on b and c quarks are consistent with a non-perturbative fragmentation
where the LQ-hadron retains an average fraction
\begin{equation}
z \approx \frac{ M_{\mrm{LQ}} }{ M_{\mrm{LQ}} + 1~\mrm{GeV} } \approx 0.995
\end{equation}
of the original LQ momentum, cf. \cite{Bjorken}. The antiquark (diquark) contributes
a typical fragmentation $p_{\perp}$ to the net $p_{\perp}$ of the LQ-hadron.
The LQ has some Fermi motion inside the LQ-hadron, of the order of the antiquark 
constituent mass, that shifts the momentum of the decaying LQ. The net effects
of all the hadronization contributions should be a more-or-less random momentum
shift at or below the 1~GeV scale, i.e. negligible. A consequence of confinement 
is that the LQ mass cannot be defined unambiguously, so the detailed effects of 
hadronization have to be seen in the context of some specific scheme for relating 
the LQ-hadron mass to the LQ mass itself.

The hadronic system will show some effects of the LQ lifetime. If an LQ-hadron 
is formed, the hadrons produced from the proton remnant and the 
initial-state-radiation partons decouple completely from those produced in
the decay of the LQ-hadron. This means, e.g., that the charged multiplicity
is independent of the LQ decay angle, i.e. of $Q^2$ (apart from some small
trigger-bias effects). By contrast, if the LQ is too short-lived for LQ-hadron
formation, the initial- and final-state partons are connected into a 
common colour string \cite{Lundmod}. The string length and hence the multiplicity
now depends on the angle between the proton remnant and the LQ hadronic decay
products (just like in standard DIS processes, with $W^2 \propto Q^2$ for fixed
$x_{\mrm{Bj}}$). If the initial- and final-state shower effects are neglected, the
variation at small $Q^2$ is significant, but is reduced to $\sim 10$\% in the
$Q^2 > 15000$~GeV$^2$ region. When showers are included, however, these tend to
dominate the particle production characteristics in a global sense. A memory 
remains in the soft (low-momentum) region, e.g. defined by $|\bf{p}| < 1$~GeV
in the longitudinal rest frame of the LQ. (The lab frame is unsuitable, since
the LQ motion in this frame introduces large spurious effects.) Over the standard
high-$Q^2$ range, and with full inclusion of parton showers, the variation in
charged multiplicity is here over 20\%. In the unlikely event that the LQ is 
very short-lived, coherence effects will also appear in the parton-shower stage, 
and so the multiplicity variation will spread to all 
$|\mathbf{p}| \lessim \Gamma_{\mrm{LQ}}$.
It is therefore possible to conclude from the hadronic final state whether an 
LQ-hadron is being formed or not, given enough statistics. This situation is quite 
similar to the one for the top quark \cite{topdecay}.     

\begin{figure}[tp]
\begin{center}
\mbox{\epsfig{file=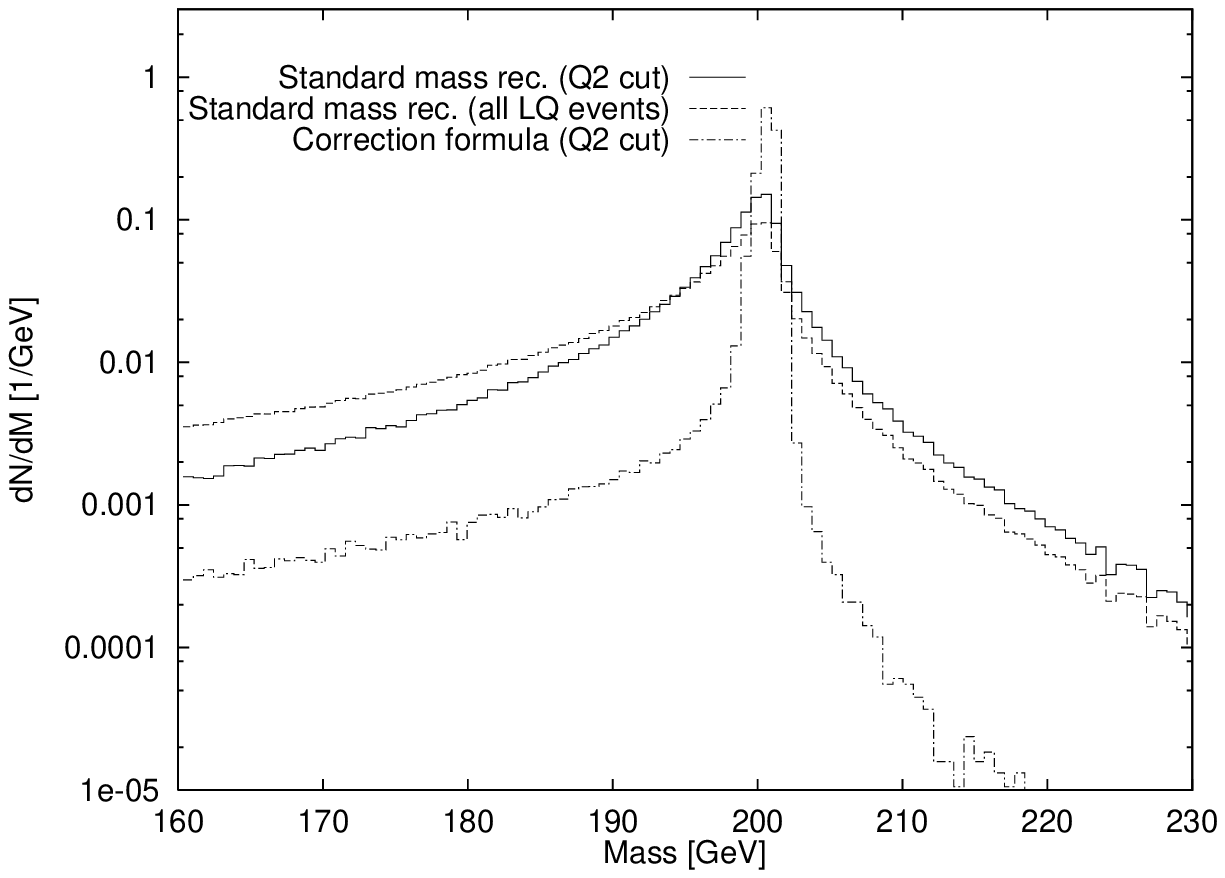}} 
\end{center}
\captive{The reconstructed LQ mass distribution 
for an input mass of 200 GeV: all LQ events with 
$m_{\mrm{recon}} = \sqrt{x_{\mrm{Bj}} s}$ (dashed), 
$Q^2 > 15000$~GeV$^2$ with $m_{\mrm{recon}} = \sqrt{x_{\mrm{Bj}} s}$
(full) and $Q^2 > 15000$~GeV$^2$ with
$m_{\mrm{recon}} = \sqrt{\tau s}$, eq. (\ref{totalshift}) (dot-dashed).
\label{mlqreconsum}}
\end{figure}

All the physics aspects described above are included in the {\sc Pythia}
event generator \cite{Manual}.\footnote{LQ-hadron formation is not part of the
standard distribution. Note that {\tt MSTJ(50)=0} and {\tt MSTP(67)=0} should be 
set to switch off interference between the initial- and final-state radiation.}
The net distribution of reconstructed LQ masses is shown in Fig.~\ref{mlqreconsum},
where all effects described above have been included. The bulk of the mass
spread is given by the effects in eq.~(\ref{totalshift}). After correction with 
this formula, the remaining mass spread is the result of the $k_{\perp}$ kick and 
virtuality of the incoming lepton, the QED radiation off the outgoing lepton,
and LQ-hadron formation. 

A $\qbar\overline{\ell}$ LQ would share most of the features described above.
There is enhanced initial-state QCD radiation in those events where a branching 
chain $\q \to \q\g, \g \to \q\qbar$ occurs. However, most events start with a
$\qbar$ at the cut-off scale $Q_0 \approx 1$~GeV, and here gluon radiation off 
the $\qbar$ is suppressed by the steeply falling $\overline{q}(x)$ distribution. 
That is, parton-shower histories that allow a $\qbar$ to survive at large $x$ and 
large $Q^2$ are biased towards lower activity (a bit more than those of the $\q$), 
and so the $\langle Q_{\qbar} \rangle \approx 12.4$~GeV rather than the 14.3~GeV of 
the $\q$. For a leptogluon, an increased radiation in the final state follows 
directly from the higher colour charge, $\langle M_{\g} \rangle \approx 45$~GeV.
In the initial state, there is again a suppression from the steeply falling gluon
distribution. About half of the events contain a branching step $\q \to \q \g$,
and this branching is biased towards higher virtualities since the quark line 
below the branching scale radiates less than the gluon above. The net result is 
a fairly large mean, $\langle Q_{\g} \rangle \approx 31$~GeV.
A leptogluon would hence give a higher hadronic multiplicity than an LQ.

So far we have concentrated exclusively on one method for LQ mass reconstruction,
based on the measurement of the scattered positron ($Q^2$, $x_{\mrm{Bj}}$ and $y$). 
Alternative methods are used by the H1 and ZEUS collaborations. However, they all 
share some common assumptions, such as massless partons and $p_{\perp}$ balance 
between the scattered positron and the recoiling hadronic system, and therefore 
do not recover the bulk of the smearing effects noted above. The optimal way to 
reconstruct a narrow(er) mass peak clearly is detector-dependent, and so must be 
worked out by the experimental collaborations. We here want to give some very 
general comments, however.

\begin{figure}[tp]
\begin{center}
\mbox{\epsfig{file=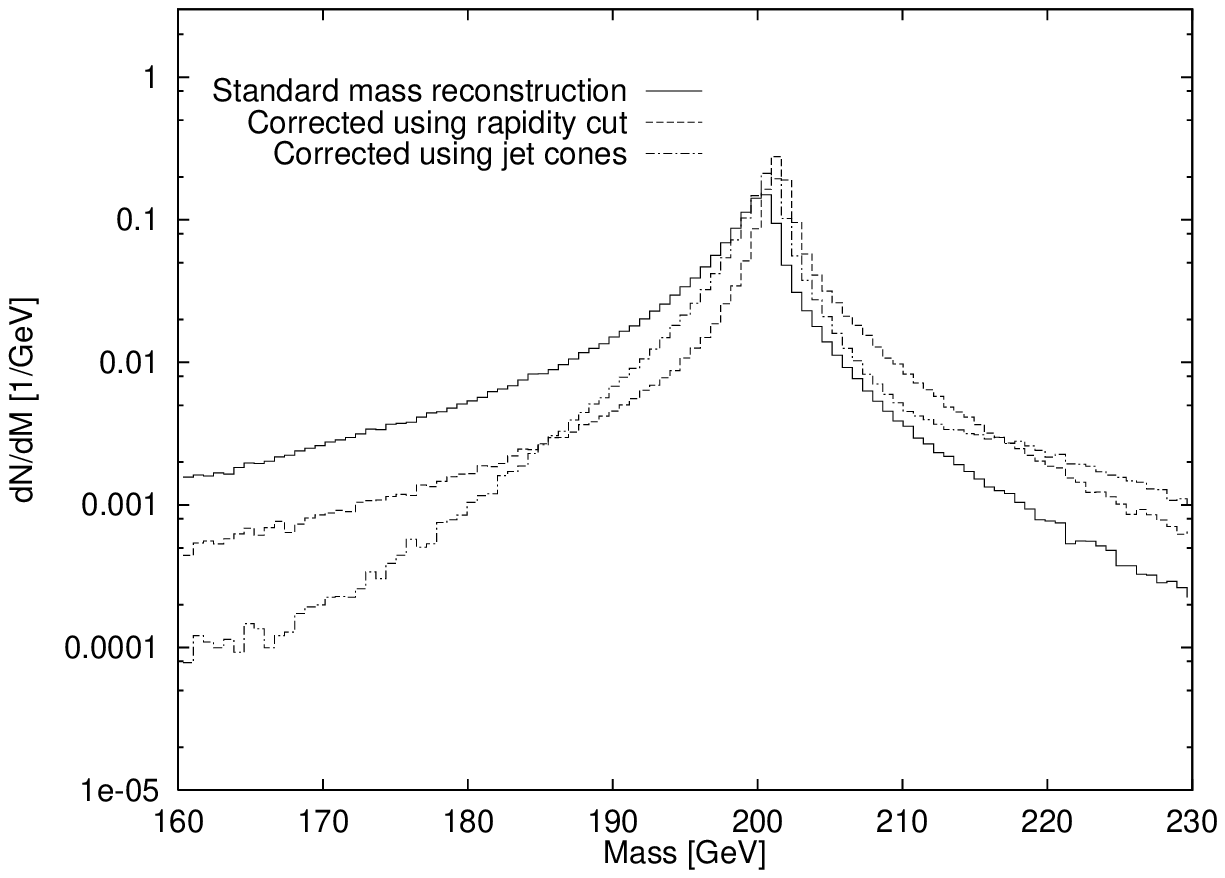}} 
\end{center}
\captive{The reconstructed LQ mass distribution for an input mass of 
200~GeV and with $Q^2 > 15000$~GeV$^2$. Curves show mass based on
$x_{\mrm{Bj}}$ only (full), or corrected for $M_{\q}$ either with a 
pseudorapidity separation (dashed) or a jet-finding strategy 
(dot-dashed). For details see text.
\label{mlqjetfind}}
\end{figure}

With an ideal detector, and good separation between the LQ decay products and the 
hadrons from the proton beam remnant, the four-momentum of the LQ could be
reconstructed to give the LQ mass directly. Such an approach suffers from the
calorimetric smearing of hadronic momentum measurements. Since the parton-shower
activity spreads out hadrons, there is also no clean separation of hadrons.
Therefore, alternatively, one could start from the better-measured positron momentum 
and then add corrections according to eq.~(\ref{totalshift}), to give a mass estimate 
$m_{\mrm{recon}} = \sqrt{\tau s}$. In practice, some of the required terms turn out 
to be rather difficult to reconstruct from the observable hadronic final state.
The most feasible correction is for the final-state QCD radiation, which we have 
seen to be the largest individual souce of smearing, and which also tends to give 
a systematic bias in a mass determination. Two alternative methods have been compared 
for reconstructing $M_{\q}$. In the first, all particles with a (pseudo)rapidity
below 1.7 (except the lepton) are used to calculate an invariant mass that is 
associated with $M_{\q}$. The cut is selected to minimize the misidentification 
of particles between the LQ-hadron and the remnant system. In the second method,
a jet clustering algorithm \cite{Manual} is used to find jets with an 
$E_{\perp} \geq 10$~GeV inside an $R = \sqrt{(\Delta\eta)^2+(\Delta\varphi)^2} < 1$
cone, and an invariant mass is calculated for the particles belonging to the 
jet or jets. The jet threshold is here selected so that initial-state radiation
is less likely to give rise to a jet. The reconstructed $M_{\q}$ is then used in
eq.~(\ref{timelikeshift}) to give a corrected LQ mass estimate. The distributions
in Fig.~\ref{mlqjetfind} show that either correction factor does result in a narrower
mass peak (the width is reduced from 11.1 to 8.2~GeV) and one that contains less 
systematic bias towards underestimating the correct mass (the average is shifted 
from 195.3 to 200.6~GeV). 
 
In summary, we have shown that several QCD effects are at play in the production 
and decay of a leptoquark. These effects are fairly well understood from our
experience in other areas of high-energy physics. It is therefore possible to 
describe in detail the production characteristics. This knowledge may be useful 
to obtain less biased and narrower mass peaks, and also to devise other tests that
could help to distinguish between leptoquark production and ordinary DIS phenomena.

\clearpage

\end{document}